\documentclass{eas}
\usepackage{graphicx}
\newcommand{\kms}          {\mbox{${\rm km~s^{-1}}$}}

\newcommand{\cc}           {\mbox{${\rm cm^{-3}}$}}

\newcommand{\ee}           {\mbox{$^{-2}$}}

\def\cm2{\mbox{${\rm cm^{-2}}$}}
\def\cm3{\mbox{${\rm cm^{-3}}$}}
\def\H2{\mbox{${\rm H}_2$}}
\def\nH2{\mbox{$n_{\rm H_2}$}}
\def\NH2{\mbox{$N_{{\rm H}_2}$}}
\def\MH2{\mbox{$M_{{\rm H}_2}$}}

\def\Msun{\mbox{$M_\odot$}}

\def\simgt{\lower.5ex\hbox{$\; \buildrel > \over \sim \;$}}
\def\simlt{\lower.5ex\hbox{$\; \buildrel < \over \sim \;$}}

\def\fe60{\mbox{$^{60}$Fe}}
\def\al26{\mbox{$^{26}$Al}}
\TitreGlobal{Physics and Astrophysics of Planetary Systems}
\begin{document}

\title{Telescopes versus Microscopes: the puzzle of iron-60}
\runningtitle{The puzzle of $^{60}$Fe}

\author{Jonathan Williams}\address{Institute for Astronomy, University of Hawaii, Honolulu, USA; jpw@ifa.hawaii.edu}

\begin{abstract}
The discovery that the short-lived radionucleide \fe60\
was present in the oldest meteorites suggests that the formation of
the Earth closely followed the death of a massive star.
I discuss three astrophysical origins: winds from an AGB star,
injection of supernova ejecta into circumstellar disks,
and induced star formation on the boundaries of HII regions.
I show that the first two fail to match the solar system \fe60\
abundance in the vast majority of star forming systems.
The cores and pillars on the edges of HII regions are
spectacular but rare sites of star formation
and larger clumps with masses $10^{3-4}$\,\Msun\ at tens
of parsec from a supernova are a more likely birth
environment for our Sun.
I also examine $\gamma$-ray observations of \fe60\ decay
and show that the Galactic background could account for the low end
of the range of meteoritic measurements if the massive star
formation rate was at least a factor of 2 higher 4.6\,Gyr ago.
\end{abstract}

\maketitle

\section{Introduction}
The study of planet formation can be approached from two sides:
the large scale encompassing the molecular core that collapses
to a star and surrounding planetary disk,
and the small scale in which the planets and the remnants of their
formation, meteorites, are examined in detail to learn about the
conditions of the early solar system (ESS).

Astronomers have identified each of the main stages by which
the interstellar medium (ISM) becomes molecular, fragments,
and individual cores collapse to protostars. We have imaged the
disks of planet forming material around young stars and
witnessed the debris from planetesimal collisions.
Although many details remain to be worked out,
the basic properties of a proto-planetary disk,
such as mass, size, composition, and lifetime, are well characterized.
We have also identified over 200 extrasolar planetary systems around
nearby stars.
The surprising diversity of these systems, however, raises the vexing
issue of how typical is our solar system.

Cosmochemists have identified the oldest rocks in the solar system
and determined their age, 4.567\,Gyr, to astonishing precision.
Careful study of their mineralogy shows the detailed conditions,
including, for example, thermal history, radiation field, and
transport processes, in the ESS. As with cosmologists
studying the Universe, however, there is one and only one system at hand
which raises the vexing issue of how typical is our solar system.

The areas where these two different lines of inquiry, telescopic and
microscopic, overlap provide interesting comparisons.
Short lived radionucleides (SLR), with half-lives less than 3\,Myr (see below),
are of particular interest for understanding planet formation and
the astrophysical environment of the ESS.
In this chapter, I focus on the puzzle presented by the presence of
\fe60\ in the most primitive meteorites.
As I show, it may be hard to reconcile the astronomical and cosmochemical
pictures but it is important to try since it may show the limitations
in our knowledge or understanding and it may also help show whether our
solar system is indeed typical, or not.

I begin by reviewing the astronomical and cosmochemical
background to this subject. These sections are brief as
they are covered in other chapters in this book.
I discuss three different scenarios which have been proposed
for the delivery of SLR into the ESS and assess the
probability for incorporation of \fe60.
I conclude that the only viable mechanism that might apply to a
large number of planetary systems is the rapid collapse of
large cluster forming clumps neighboring a massive star forming region.
Finally I return to a basic assumption that the \fe60\ abundance is
greater than the Galactic background and show a discrepancy between
the predicted level and recent $\gamma$-ray observations.

\section{Astronomical observations of star and planet formation}
Stars form in molecular clouds, strung out along the spiral arms
of the Galaxy. Our current understanding of the processes by which
these large and massive clouds condense to stellar
scales was recently summarized by McKee \& Ostriker (\cite{McKee07}).
A rotationally supported disk inevitably accompanies the protostar
due to conservation of angular momentum and the magnification
of any initial spin. These disks initially funnel material onto
the growing star but subsequently become the sites of planet formation.
The observational properties and the theory behind the
coagulation of sub-micron sized ISM dust grains to planetesimals
are reviewed in the chapters by Hogerheijde and Youdin.

For the purposes of the discussion here it is important to note
that several states, or phases, of gas coexist in the ISM (Cox \cite{Cox05}).
The lowest density state is a hot ionized phase produced by supernovae
and this fills most of the volume. Most of the mass is in atomic
clouds at intermediate temperatures.
The highest density state is cold and molecular.
Most stars form in giant molecular clouds with typical
sizes\footnotemark\footnotetext{I use $\sim$ to indicate an order
of magnitude variation around a value and $\approx$ for a factor of 3.},
$L\approx 50$\,pc and masses, $M\sim 10^5$\,\Msun.
The clouds possess considerable substructure, generically
characterized as clumps with sizes $L\approx 5$\,pc and masses,
$M\sim 10^3$\,\Msun\ (Williams, Blitz, \& McKee \cite{Williams00}).
Individual stars form in dense cores within clumps, with typical radii
$R\approx 0.05$\,pc and mass $M\approx 1$\,\Msun\
(di Francesco et al. \cite{diFrancesco07}).
Protostellar disks have radii $R\approx 100$\,AU and masses ranging from
$M\approx 10^{-3}-10^{-1}$\,\Msun\ (Andrews \& Williams \cite{Andrews07}).

Although many observational and theoretical studies are directed toward
the formation of isolated, individual stars, as this is the simplest
system to understand, most stars form in large clusters
(Lada \& Lada \cite{Lada03}). The stellar mass distribution is heavily
skewed toward the low end in terms of numbers but the few massive
stars dominate the luminosity and their radiation, winds, and
eventual supernovae may significantly affect the properties of
neighboring primordial planetary systems.

The timescales are also critical for our comparison with cosmochemistry.
It is surprisingly difficult to identify young or old clouds
(or even tell the difference -- see Williams \& Maddalena \cite{Williams96})
and consequently the lifetimes of giant molecular cloud are highly uncertain.
Current estimates, based on theories of cloud formation and destruction,
range from $\approx 1-20$\,Myr
(Hartmann, Ballesteros-Paredes, \& Bergin \cite{Hartmann01};
Matzner \cite{Matzner02}).
The evolutionary state of cores within star forming clouds
are easier to characterize and several lines of evidence suggest that
as soon as they become sufficiently dense,
$\nH2\sim 10^5$\,\cc, stars will rapidly form within a few
free-fall timescales, $\sim 10^5$\,yr (Jorgensen et al. \cite{Jorgensen07}).
The presence of a circumstellar disk around a star is most readily
detected as a long wavelength excess above the stellar photosphere.
Large surveys of clusters with different ages
show that the fraction of stars with disks decreases from an initial
value close to unity to zero by by 6\,Myr, and that the median disk lifetime
is about 3\,Myr (Haisch, Lada, \& Lada \cite{Haisch01};
Hernandez et al. \cite{Hernandez08}).
This is the characteristic timescale for the formation of planetesimals
although the last steps in the growth of fully fledged planets
may take considerably longer (see chapters by Kalas and Beichmann).

\section{Short lived radionucleides and the importance of $^{60}$Fe}
Astronomers are limited to studying the light that molecular cores,
young stars, and dusty disks either emit naturally or absorb and scatter
from a background or nearby source. Having rocks in the laboratory, however,
allows for far more targeted and detailed investigations. Cosmochemists
can break down individual grains within meteorites to the atomic
level and measure their structure, mineralogy, and isotopic
composition to learn about the precise conditions in which they formed.

An element can have unstable isotopes that are chemically identical but
decay over cosmic timescales. A dust grain with a particular mineralogy
can therefore be incorporated into a planetesimal but subsequently change
its composition. These isotopic anomalies are then frozen into the material
and, unless the planetesimal undergoes further processing, are immutable.
SLR, defined here as having half-lives less than the 3\,Myr
characteristic lifetime of disks, therefore provide a natural chronometer
for planet formation.
I refer the reader to the chapters by Aleon and Gounelle
for more detail and the broader aspects of such work.

\fe60\ decays to $^{60}$Ni with a half life of 1.5\,Myr
and its presence in the ESS was inferred by the
correlation\footnotemark\footnotetext{~If \fe60\ had not been present,
the nickel ratio would have been independent of iron.}
of the isotope ratio, $^{60}$Ni/$^{61}$Ni,
with the main isotope of iron, $^{56}$Fe/$^{61}$Ni
(Tachibana \& Huss \cite{Tachibana03}).
The exact abundance of \fe60\ in the ESS is not yet firmly established,
with published measurements ranging from
[\fe60]/[$^{56}$Fe]$=3-10\times 10^{-7}$
(Tachibana et al. \cite{Tachibana06}; Moynier et al. \cite{Moynier05})
but it appears to be greater than the expected background level.

The Galactic background is set by a balance between production in massive
stellar winds and supernovae and destruction by radioactive decay.
The equilibrium abundance of
any radionucleide, normalized by the ratio of production rates, is then a
simple fuction of its decay time (Schramm \& Wasserburg \cite{Schramm70}).
Jacobsen (\cite{Jacobsen05}) plots the normalized abundances of many
radionucleides with half-lives from 0.1\,Myr to almost 10\,Gyr against
the expected background. Solar system levels are generally lower than
the ISM average but most radionucleides with half-lives greater than
3\,Myr fit a model with a slow exchange, over a $\approx 60$\,Myr timescale,
between the hot ionized medium and cold, molecular clouds.
The background level decreases rapidly with decay time and many
SLR lie above this model, however.

The origin of radionucleides with abundances above the background
can be broadly categorized in two ways:
production by energetic particles from the active protosun
and rapid transport from an external source into the ESS
(Wadhwa et al. \cite{Wadhwa07}).
Both mechanisms have considerable flexibility and can be adjusted to
match the abundances of most SLRs. Note that these are not mutually
exclusive and both may have played a role.
For instance, $^{10}$Be can only be produced by spallation
reactions, and is strong evidence for disk processing by protosolar radiation.
On the other hand, the neutron rich iron isotope, \fe60,
can only be formed in the cores of massive stars and must have
then been delivered into the ESS on a timescale comparable to
its half life.

The possibility of an external influence on the Sun's formation
predates the discovery of \fe60\ (see Cameron \& Truran \cite{Cameron77}
and references therein) but the success of local irradiation models
(e.g. Lee et al. \cite{Lee98}; see also the chapter by Gounelle
in this volume) allowed most astronomers to downplay the issue.
The importance of \fe60\ is that it is most clear-cut example of
an SLR that could not have been produced by the protosun and therefore
constitutes indisputable evidence that the products of nucleosynthesis
from the core of massive star polluted the protosolar nebula.
The puzzle of \fe60\ is that astrophysical contexts in which
this occurs do not appear to be common. We are faced with either a
significant gap in our understanding of star and planet formation or the
realization that our solar system formed in a very unusual situation.

\section{Possible origins of $^{60}$Fe in the early solar system}
\subsection{Pollution of the protosolar nebula by an AGB star}
Asymptotic Giant Branch (AGB) stars are evolved low and intermediate
mass stars that have used up all the hydrogen in the central core
and are now burning hydrogen and helium in shells around it.
Mixing in these shells dredges up newly synthesized material
from the core which may be blown out in a powerful stellar wind.
In principle, SLRs with abundances comparable to ESS levels can
be injected into the ISM through AGB winds (Cameron \cite{Cameron93}).
Figure~1 shows a wind-blown bubble from an evolved massive star in a
star-forming molecular cloud; a simultaneous demonstration that this
situation can occur but that its effect is limited to a very small
region of the cloud.

In regard to the \fe60\ problem discussed here,
its production in an AGB star requires faster reactions than
that required for \al26. Wasserburg et al. (\cite{Wasserburg06}) show that
only stars with masses $\simgt 5$\,\Msun\ produce enough \fe60\ to
match the ESS abundance. Stars with masses $>10$\,\Msun\ end their
lives as supernovae and the possibility that they are the origin of
the \fe60\ is discussed below. The stellar mass range of interest
for the AGB hypothesis is therefore $5-10$\,\Msun.

By their nature, AGB stars
have ages a little greater than the main sequence lifetime appropriate
to their mass, $\sim 25-110$\,Myr (Schaller et al. \cite{Schaller92}).
The molecular cloud in which they were born will have dispersed
and any connection to ongoing star and planet formation will be
serendipitous. Based on a census of the known mass-losing
AGB stars and molecular clouds within 1\,kpc of the sun,
Kastner \& Myers (\cite{Kastner94}) estimated the probability
of a chance encounter to be $\sim 1$\% per Myr.
However, the ratio of AGB wind to molecular mass is very small,
$3\times 10^{-4}$, and they concluded -- in the context of \al26\ -- that the
probability that the ESS was polluted in this way was $\le 3\times 10^{-6}$.

\begin{figure}[ht]
\includegraphics[height=12.2cm,angle=90]{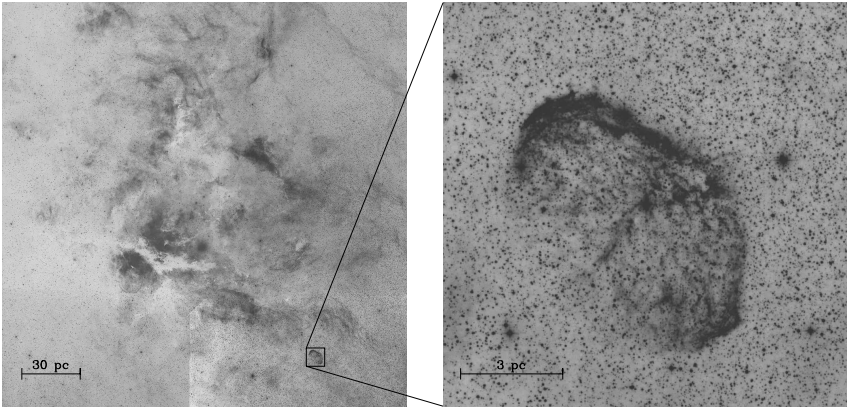}
\caption{The relative scale of AGB ejecta compared to molecular clouds.
The two panels are from the Digitized Sky Survey and show optical
extinction and nebulosity associated with the IC1318 star forming
region in the left panel. The right panel shows a close-up of the
wind-blown bubble from the AGB star, HD\,192163. Note the relative
scales that graphically demonstrates how small a part of a molecular
cloud can be polluted by a massive star wind.}
\end{figure}

The Kastner \& Myers work is an often cited argument against an
AGB stellar origin for \al26\ in the ESS. The same reasoning applies
to \fe60\ with an even lower likelihood due to the higher stellar
masses necessary. Their result was based on the catalogs of AGB
stars and molecular clouds in the solar neighborhood (1\,kpc), however,
and it is worth making a related but more general case for the
entire Galaxy.

Jura \& Kleinmann (\cite{Jura90}) find that the average surface
density of AGB stars with high mass loss rates is $\sim 10$\,kpc\ee,
independent of Galactic radius. This implies a total population
of about $3\times 10^3$ in the Galaxy. Each lasts for about 1\,Myr,
a timescale comparable to the \fe60\ half-life, and ejects a total of
about 3\,\Msun\ into the ISM.
Based on the predicted wind SLR abundances,
these winds can pollute about 100 times more mass to ESS levels
or $\sim 10^6$\,\Msun\ in total. The total molecular mass in the Galaxy is
$1\times 10^9$\,\Msun\ (Williams \& McKee \cite{Williams97}) so the
proportion of star forming material that can be enriched is no more than 0.1\%.
This assumes that the star formation efficiency is independent
of the presence of an AGB star which is entirely consistent with observations.
Note that this is a generous upper limit since it does not
include the low likelihood that an AGB star lie within or
near a molecular cloud as in Kastner \& Myers, nor does it
include SLR decay during the (potentially $\gg 1$\,Myr)
passage from AGB wind to planetesimal. Finally the Jura \& Kleinmann
AGB number count likely includes many stars with masses too
low to produce appreciable \fe60. Even without these additional factors,
however, this simple calculation provides a robust and independent
demonstration that the pollution of the proto-solar system
by the winds from an AGB star is a very unlikely scenario for the origin
of \fe60\ in the ESS.

\subsection{Injection of supernova ejecta into the protoplanetary disk}
Circumstellar disks can survive a supernova explosion as close as 0.2\,pc
and potentially capture significant levels of SLR from the ejecta
(Chevalier \cite{Chevalier00}). Subsequent numerical simulations by
Ouellette, Desch, \& Hester (\cite{Ouellette07}) show that solid grains
can be efficiently mixed into the disk material. The mechanism works but
how often does it occur?

A fundamental constraint is the similarity in the timescales for massive star
and disk evolution (Figure~2). Even the most massive stars take 3\,Myr
to burn their hydrogen and evolve off the main sequence, at which point
half of the disks around neighboring stars have disappeared. Further,
only stars with masses greater than 30\,\Msun\ explode within 6\,Myr,
the maximum disk lifetime. Stars this massive are extremely rare and only
found in the largest clusters.

\begin{figure}[ht]
\includegraphics[height=12.2cm,angle=90]{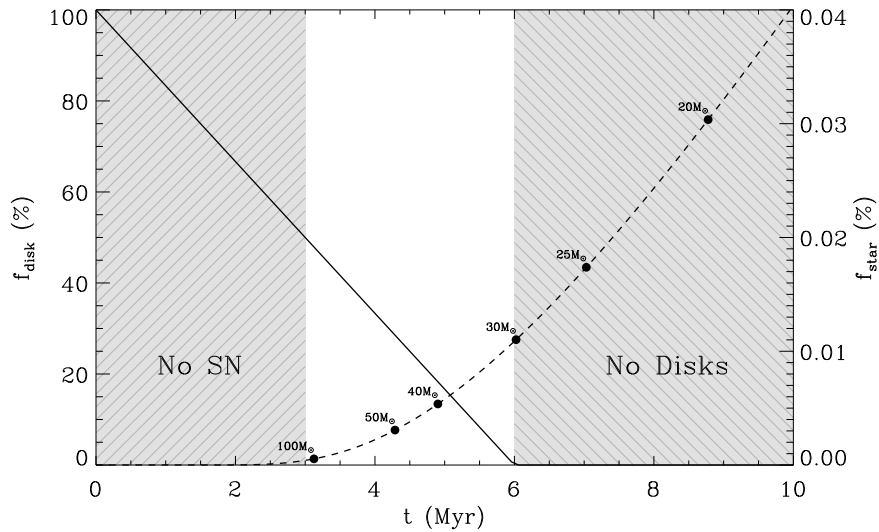}
\caption{Disk lifetimes versus main-sequence stellar timescales.
The solid diagonal line decreasing from 100\% to 0\%
represents the decrease in the disk fraction in clusters of
varying age and is taken from Haisch et al. (\cite{Haisch01}).
The curved dashed line plots the percentage of stars with
main sequence lifetimes from 3 to 10\,Myr. The hashed areas
show that there are no supernovae before 3\,Myr and no disks
remain after 6\,Myr. This leaves only a small range in time,
$3-6$\,Myr, when there are both disks and a supernova.
Moreover, only stars with masses greater than
30\,Msun\ satisfy this constraint and they are extremely rare,
numbering only about one in $10^4$ stars.}
\end{figure}

An even stronger constraint comes matching the abundances of the
supernova ejecta with the meteoritic record.
The amount of captured material is equal to the product of the disk area
and the surface density of the supernova ejecta. The abundance of
a particular SLR is therefore proportional to the supernova yield
times the square of the ratio of the disk radius to the
distance from the source. Ouellette et al. (\cite{Ouellette07}) and
Looney, Tobin, \& Fields (\cite{Looney06}) show this
criterion requires the disk be within a ``radioactivity distance''
of 0.3\,pc of a supernova with progenitor mass $10-40$\,Msun.
This is only slightly greater than the survival distance.
The disk injection scenario therefore only works over a very
narrow range of distances and this makes it extremely unlikely.

In Williams \& Gaidos (\cite{Williams07}), we explicitly calculate the disk
injection likelihood, taking into account the massive star
lifetimes, disk evolution, and the constraints on distance from
the source. We considered a cluster with $N_*$ stars sampled from
the initial mass function (IMF), $dN_*/dM_*\propto M_*^{-2.5}$ (Scalo 1986),
and calculate the most likely supernova progenitor mass.
This defines a supernova timescale and a corresponding disk fraction
(see Figure~2).
To take the disk-supernova distance into account, we model the
cluster as expanding linearly with time and parameterize
the expansion via the surface density at 3\,Myr,
an observationally defined quantity.
We also extended the enrichment range out to 0.4\,pc to allow for
more massive supernovae than considered by Ouellette et al. and Looney et al.

The results are shown in the left panel of Figure~3. 
The probability that a disk is injected peaks at $0.5-1.5$\%,
depending on stellar density, at $N_*\simeq 10^4$.
Small clusters are unlikely to harbor a supernova that explodes
before all the disks have disappeared and most disks are
too far from the supernova to be enriched in large clusters.
In fact, the enrichment fraction is the dominant factor
in the low overall probability; clusters with $10^4$ stars have
about about a 50\% chance of containing a high mass star that will
become a supernova within 4\,Myr implying a disk fraction of 1/3 at the
time of the blast, but the cluster will be several
parsecs in radius and only a few percent of disks will be
close enough to the supernova to be enriched.

\begin{figure}[ht]
\includegraphics[width=5cm,angle=90]{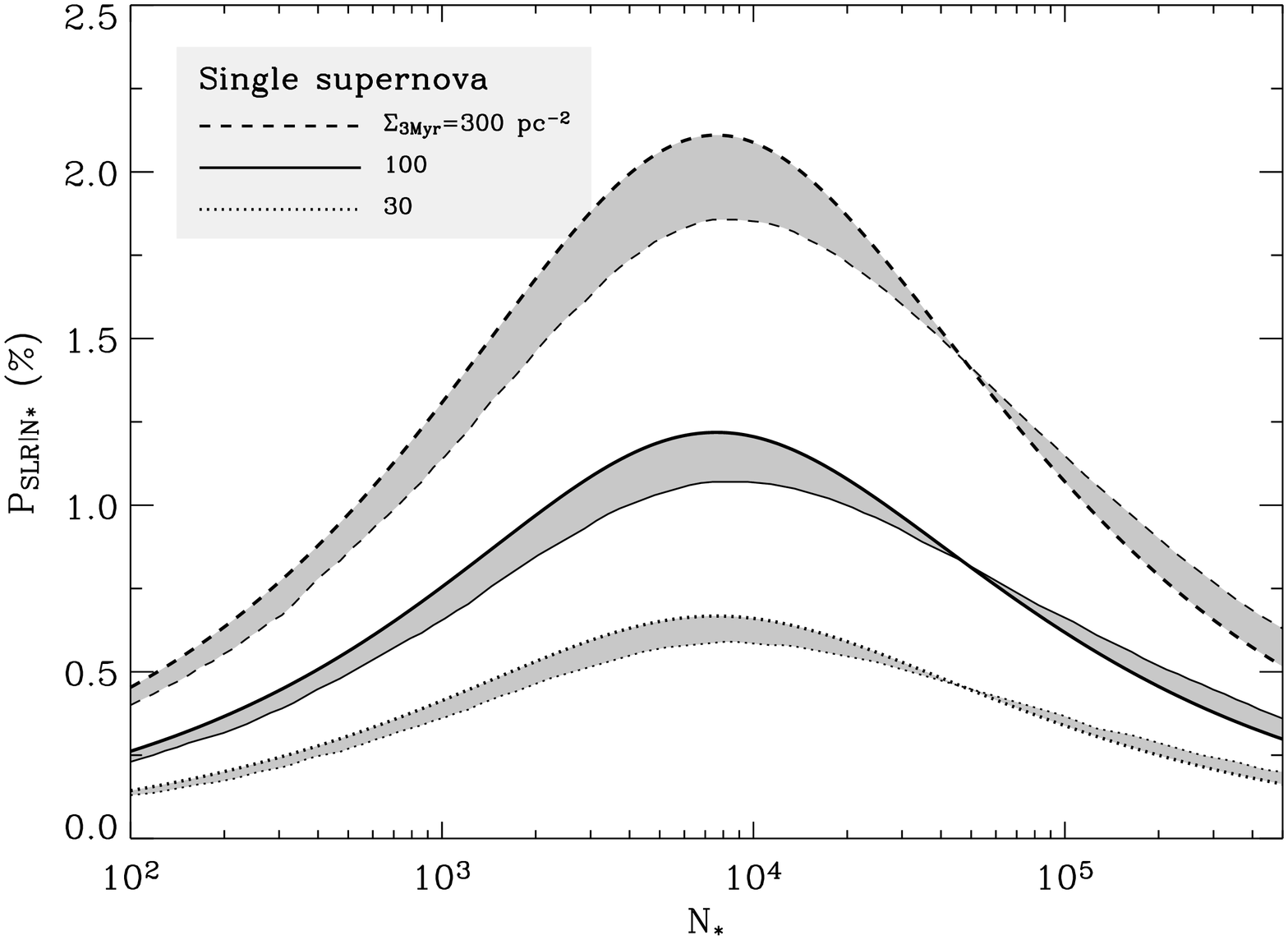}
\includegraphics[width=5cm,angle=90]{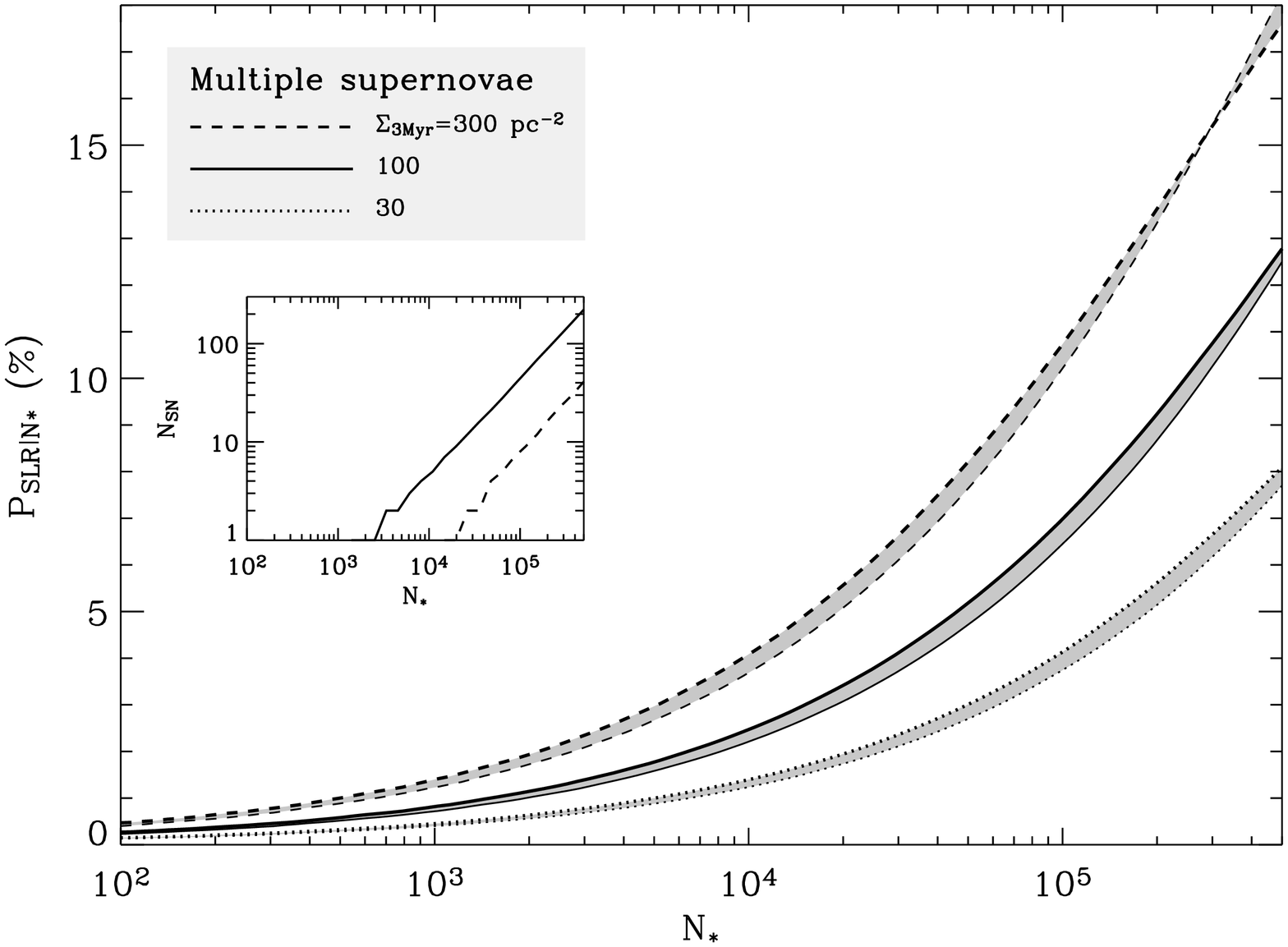}
\caption{Probability of enrichment of a protostellar disk
by a supernova versus cluster size. In each panel three curves
are plotted corresponding to different expansion rates,
parameterized by the stellar number density, $\Sigma_{3Myr}$.
The shading shows the range of probabilities over cluster formation
timescales from 0 to 3\,Myr. The left panel shows the disk enrichment
likelihood for a single supernova event. Small clusters are unlikely
to have a supernova before all disks have dissipated and most disks
in large clusters are too far away from the supernova to be enriched
to ESS levels. The right panel shows the disk enrichment probability
in the case where multiple supernovae can enrich multiple disks.
In this case, the probability rises strongly in large clusters
where many supernovae occur (see inset) and many disks may be impacted.}
\end{figure}

It is also possible to relax an implicit assumption in the model
that all the stars in the cluster form at once. This does not
greatly change the disk injection probability. The greyscale
associated with each surface density in Figure~3 shows the range
of probabilities for a cluster formation timescale varying from
$0-3$\,Myr. Except for the largest clusters, $N_*\simgt 10^5$ stars,
the probability decreases with formation time since the rarer, massive
stars will likely form after many low mass stars and their disks will
have an evolutionary head start. This is for an unbiased sampling of the
IMF. The probability can be increased by putting in a bias toward
forming high mass stars first (equivalent to shifting the stellar
curve leftward in Figure~2, but there is no observational evidence
for this. Young high mass stars are always surrounded by many low
mass stars and no O star has been observed in isolation at less
than 1\,Myr.

Large clusters can host many supernovae and these can enrich multiple disks.
In this case, and generously assuming that the low and high mass stars are
spatially mixed\footnotemark\footnotetext{~In practice,
high mass stars tend to lie together at the cluster center and their
combined impact on the other stars is reduced.},
the right panel in Figure~3 shows that
the disk injection probability rises substantially to $\approx 10$\%
in the largest known clusters, $N_*=5\times 10^5$.
However, most stars form in smaller clusters
(actually an equal number per logarithmic bin) and the overall
disk injection probability, for any given star in the Galaxy,
is very small, $\simlt 1$\%. Nevertheless, this shows that if
the proto-solar system disk was impacted by a supernova blast, it most
likely happened in an enormous cluster, similar in size or greater than
W49 or $\eta$ Carinae today.

This simple model can be expanded in a number of ways but it is hard
to escape the fact that very few disks are enriched with SLR by direct
incorporation of supernova ejecta.
Gounelle \& Meibom (\cite{Gounelle08}) include
an additional factor to allow for external photoevaporation of the
disk by massive stars in the above formalism. This reduces the
disk injection probability further, particularly in the most
massive clusters where it negates the effect of multiple supernovae.
Certainly massive stars can rapidly erode the outer parts of disks,
as observed in the Trapezium Cluster in Orion (O'dell \& Wen \cite{Odell93})
but it is not yet clear that this prevents planet formation in
the inner parts. Indeed Throop \& Bally (\cite{Throop05}) postulate
that the preferential removal of gas might enhance dust sedimentation
and speed up the growth of planetesimals.
Williams, Andrews, \& Wilner (\cite{Williams05}) show that enough mass
remains in several Trapezium Cluster disks to form solar system
scale architectures but more observations are required to show the
statistics and significance of photoevaporation on disk mass
(Mann \& Williams, in prep). On the other hand, mid-infrared observations
by Balog et al. (\cite{Balog07}) and Hernandez et al. (\cite{Hernandez08})
show marginally significant evidence for a decrease in the inner disk
fraction of about a factor of 2 within 0.5\,pc of massive stars.

Our model shows that the disk injection probability increases
for higher cluster surface densities. We bracketed a range,
$\Sigma_*\simeq 30-300$\,stars\,pc$^{-2}$, that is observed in infrared surveys
(Adams et al. \cite{Adams06}, Carpenter et al. \cite{Carpenter00}).
This is also consistent with Lada \& Lada (\cite{Lada03}) who show that
the number of detectable clusters declines with age and estimate
that only about 10\% survive as recognizable entities beyond 10\,Myr
(i.e. with surface densities significantly above the field star background).
Our Sun could not have formed in a long-lived cluster, however,
because the solar system would have been disrupted by stellar
encounters (Adams \& Laughlin \cite{Adams01}).

Finally, we have treated supernova ejecta as isotropic but observations
show a great deal of inhomogeneity (e.g. Hwang et al. \cite{Hwang04}).
In principle, this allows disk enrichment at greater distances
but the overall disk injection probability decreases
because more disks nearby are not enriched.
Specifically, let the filling factor be $f$.
Then the ejecta are spread over an area $4\pi d^2f$ at distance
$d$ from the supernova. Compared to the homogeneous case, a disk can be
enriched at a greater distance, $\propto f^{-1/2}$,
but the disks are distributed isotopically and the number that are
impacted decreases linearly with $f$.
The overall number of injected disks is the product of these and therefore
proportional to $f^{1/2}<1$. Consider the extreme case where all the ejecta
are shot out in a single beam; at most one disk may be enriched
very far from the star but all the other disks in the cluster
are not enriched at all.

There does not appear to be any way to make this mechanism apply in
a general way and I therefore conclude that the direct injection of
supernova ejecta into the proto-solar system disk is an unlikely scenario
for the origin of \fe60\ in the ESS.

\subsection{Induced star formation}
The fundamental puzzle of \fe60\ is that it associates the birth of
our Sun with the death of a massive star.
Yet stellar evolutionary timescales are
generally much longer than formation timescales. Even the most
massive stars, with the shortest lives, require 3\,Myr
before becoming a supernova and ejecting SLR into their surroundings. This
is much greater than the $<1$\,Myr timescale for a dense core
to form and collapse to a protostar.

As shown above, the timescale problem is one of the factors in the
low likelihood of AGB wind contamination of the proto-solar system
or direct injection of supernova ejecta into the proto-planetary disk.
This problem would be mitigated, however, if there were a causal relation
between the death of a massive star and the birth of lower mass stars.
This led to the idea of induced star formation as a solution to
the SLR problem.

It has long been known that large stellar associations in molecular
clouds often consist of spatially and kinematically distinct subgroups
ordered in age (Blauuw 1964).
The canonical example is the four Orion groups, OB1a--d,
spaced sequentially from north to south and with ages
from $\approx 10$ to $\approx 1$\,Myr respectively.
Elmegreen \& Lada (\cite{Elmegreen77}) proposed that such
sequences could be explained by the induced collapse of molecular
gas on the boundary of an expanding HII region.

There is a large body of literature on the observations and theory
of star formation on the boundaries of HII regions
(e.g., see the recent conference proceedings
by Elmegreen \& Palous \cite{Elmegreen07})
and clearly some stars are induced.
Hester \& Desch (\cite{Hester05})
have advocated that our Sun formed in this manner and
incorporated SLR from the massive stellar winds and subsequent supernovae.
As with the other SLR transport mechanisms, however, the essential
question is not whether this might occur but how likely is it?
The answer depends on the size scale involved.

\newpage
\noindent{\itshape\sffamily Core scales, $L<0.1$\,pc}

The expansion and photoionization of an HII region sweeps away the
inter-clump material in a molecular cloud and reveals the denser regions.
Indeed part of the reason why images of star formation on the
boundaries of HII regions are so spectacular is because fingers of dusty,
molecular gas are dramatically silhouetted against a bright background
(e.g., Hester et al. \cite{Hester96}; Smith, Stassun, \& Bally \cite{Smith05}).
However, these structures amount to only a small fraction of the total mass
of molecular gas in the cloud, are very short-lived, and they do not
appear to have a significantly enhanced star formation efficiency.
For instance, the famous ``pillars of creation'' in M16 have a total mass
$M_{\rm gas}=200$\,\Msun\ (White et al. \cite{White99})
and their high velocity gradients indicate dynamical timescales
$\sim 10^5$\,yr (Pound \cite{Pound98}).
Only 11 of the 73 globules within the pillars contain a young stellar object
(McCaughrean \& Andersen \cite{McCaughrean02}) which indicates
a star forming efficiency, $M_*/M_{\rm gas}\approx 3$\%,
similar to molecular clouds in general
and a total number of induced stars that is dwarfed by the
$\approx 10^4$ stars in the host HII region
(Hillenbrand et al. \cite{Hillenbrand93}).

Even if there were 100 such star-forming pillars at the cloud
interface over the lifetime of the HII region, the total number
of stars that were induced to form would amount to only about
10\% of the stars in the central cluster.
Obviously this is an extremely rough estimate but new high quality
infrared and optical archival datasets should allow a more precise
accounting of the amount of stars that are induced versus those that
form spontaneously, and those in the associated HII regions.

As with the other SLR transport mechanisms, the geometric dilution of
SLR as they travel from source to planetesimal is the strongest constraint.
Looney et al. (\cite{Looney06}) show that the radioactivity distance
that matches the \fe60\ abundance with the ESS is about
100 times the core radius. For a typical core with radius
$\approx 0.05$\,pc this corresponds to $\approx 5$\,pc.
Note also that this assumes 100\% transport efficiency from the
supernova to the collapsing core which is likely to be overestimated
by an order of magnitude (Vanhala \& Boss \cite{Vanhala02}).
A 10\% efficiency would imply a supernova-core distance of $\approx 2$\,pc.

The M16 pillars are about 2\,pc from the ionizing star so appear to be an
ideal analog for the solar birthplace (Hester \& Desch \cite{Hester05}).
However, HII regions expand rapidly to tens of parsecs in size
and any molecular cores or circumstellar disks must survive the
direct impact of stellar wind and radiation for at least 3\,Myr
until the supernova occurs.
Numerical simulations by Mellema et al. (\cite{Mellema06})
show that most molecular globules at the center of an HII region
are photoablated within 0.4\,Myr.
Freyer, Hensler, \& Yorke (\cite{Freyer03}) include the
effect of stellar winds and show that they sweep away molecular
filaments inside the HII region to beyond 10\,pc by 3\,Myr.

The impact on a stellar wind on a core and whether it might
trigger star formation was studied by Foster \& Boss (\cite{Foster96}).
They found that the core could be induced to collapse if it remained
isothermal (i.e., radiates away the impact energy) during the
passage of the shock and that this occurs only for speeds less than
100\,\kms. Stellar winds are much faster than this and
their impact will shred a core.
A simple visual comparison of their simulations with images
of molecular globules inside HII regions look closer to the
fast shock case suggesting that the dominant effect is core
destruction and not star formation (Figure~4),

\begin{figure}[ht]
\includegraphics[height=12.2cm,angle=90]{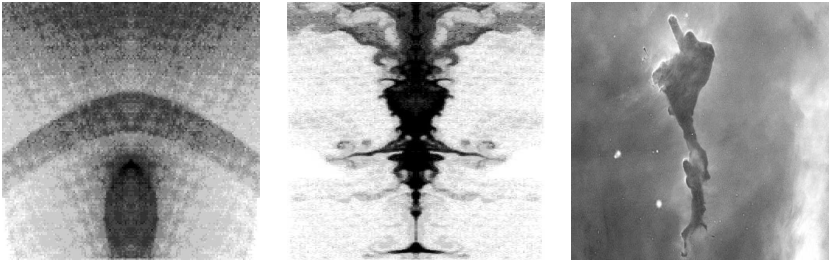}
\caption{Comparison of numerical simulations of induced core
collapse with observations. The left two panels show results
from Foster \& Boss (\cite{Foster96}) on the density distribution
of dense cores impacted by slow and fast stellar ejecta.
Speeds greater than 100\,\kms\ shred cores rather than
induce them to collapse. The right panel shows an
image of the Keyhole nebula in the $\eta$ Carina nebula.
This is a dense core that appears to be in the process of being
destroyed by the HII region rather than compressed to form new stars.}
\end{figure}

Observations of the molecular gas toward HII regions with Myr ages
often show a ring of enhanced column density on the boundary with
the cloud but very little if any molecular material in the center
(e.g., Lang et al. \cite{Lang00}). Individual stars in clusters
have an age spread of $\approx 1$\,Myr but even if some stars were
induced during the
early expansion of the HII region their surrounding core has gone
and any future supernova will, at most, impact a circumstellar disk.
That is the scenario discussed above and shown to affect less than
1\% of all protoplanetary systems.
Since the evolution of the disk fraction includes
observations of OB associations, any early induced star
formation of this nature is implicitly taken into account.

\bigskip
\noindent{\itshape\sffamily Clump and cloud scales, $L>0.1$\,pc}

Larger targets capture more ejecta and can be further from the
supernova. On the scales that characterize clumps and clouds,
the radioactivity distance ranges from
$50-10^4$\,pc and better matches the sizes of evolved HII regions.

\begin{figure}[ht]
\includegraphics[height=12.8cm,angle=90]{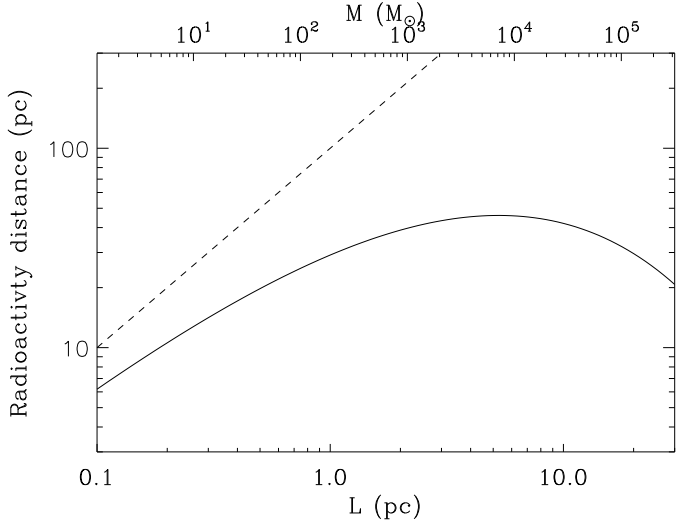}
\caption{The radioactivity distance at which an object of
size $L$ will receive \fe60\ to produce an abundance relative
to $^{56}$Fe that matches the ESS level. The dashed line shows
the Looney et al. (\cite{Looney06}) calculation with no decay
and the solid curved line incorporates decay during the free-fall collapse.}
\end{figure}

The issue here is that the large regions must collapse to stellar
scales and there will be significant decay of \fe60\ as this
happens. Larger regions have lower average densities and longer
free-fall timescales. Using the size-linewidth relation in
Heyer \& Brunt (\cite{Heyer04}), and deriving a density from
the virial mass, implies
$t_{\rm ff}{\rm (Myr)}=3.7L{\rm (pc)}^{0.41}$.
The initial amount of \fe60\ must be higher to compensate
for the decay and this reduces the clump-supernova distance
by a factor $e^{t_{\rm ff}/2t_{1/2}}$ where $t_{1/2}=1.5$\,Myr\
(Figure~5)\footnotemark\footnotetext{Note that the free-fall timescale
is much greater than the travel time for the ejecta to reach the source
and the latter is therefore neglected.}.

The longer collapse times at large size scales offsets the greater amount
of initial \fe60\ and the radioactivity distance has a maximum of 45\,pc.
Objects further away than this will be either
too small to capture enough \fe60\ or too large to collapse
before significant decay. However clumps with sizes $L=2-11$\,pc
and masses $M=10^3-4\times 10^4$\,\Msun\ that are within 40\,pc
of the massive stars and collapse as (or because) a supernova explodes
could, in principle, produce numerous planetary systems enriched
with ESS levels of \fe60. Of all the possibilities considered so
far, these objects represent the most promising candidates to match
the cosmochemical constraints on the birth environment of the solar system.

\bigskip
\noindent{\itshape\sffamily The case of the Rosette}

The Rosette molecular cloud, with mass $2\times 10^5$\,\Msun,
sits adjacent to the well known Rosette nebula,
a luminous 2--3\,Myr old HII region powered by the NGC2244 cluster
containing 7 O stars and about $10^4$ stars in total
(Townsley et al. \cite{Townsley03}).
The cloud contains several embedded clusters
(Roman-Zuniga et al. \cite{Roman08}) indicative of ongoing star formation.
Mid-infrared observations by Poulton et al. (\cite{Poulton08})
reveals circumstellar disks in the nebula in addition to protostars
in the cloud (Figure~6).

\begin{figure}[ht]
\includegraphics[height=12.2cm,angle=90]{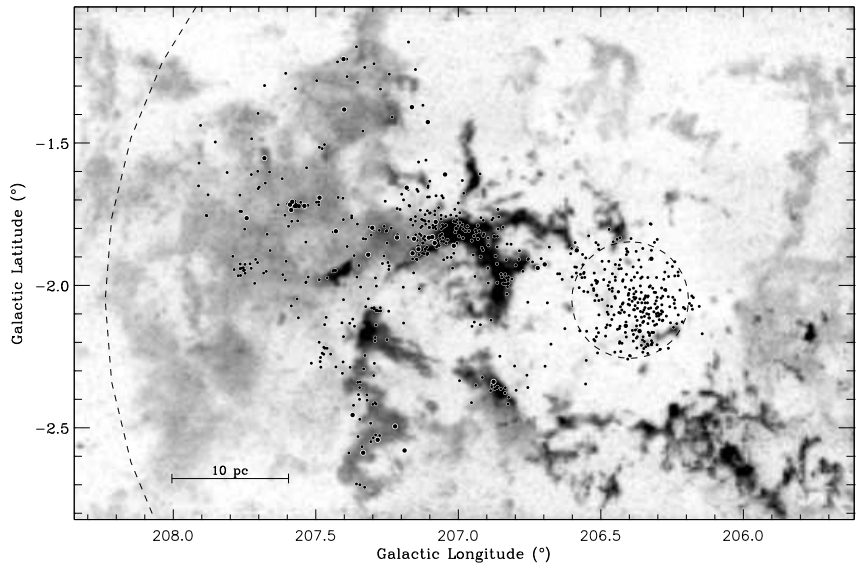}
\caption{Young stars in the Rosette molecular cloud. The greyscale
plots CO emission from the cold gas in the cloud
(Heyer, Williams, \& Brunt \cite{Heyer06}) and the dots show
the location of stars with mid-infrared excesses
(Poulton et al. \cite{Poulton08}).
These dusty sources stars form two distinct groups;
a cluster of disks within the HII region remnants of its
formation $2-3$\,Myr ago, and a distributed
population of protostellar disks and envelopes in
the cloud, concentrated toward the molecular peaks.
The dashed lines show the 5\,pc radius around the cluster
at which dense cores could be enriched (if they survived)
and the maximum 45\,pc radioactivity distance at which
clumps with size 5\,pc could be enriched.}
\end{figure}

There are 293 identified disks in the NGC2244 cluster and their
detection statistics are in proportion to that expected for its
$2-3$\,Myr age (Balog et al. \cite{Balog07}).
458 sources lie in the cloud, mostly concentrated toward molecular
peaks, and possess spectral energy distributions indicative of
protostellar disks or envelopes and ages $\approx 1$\,Myr.

The HII region has evacuated a noticeable spherical cavity
in the molecular gas with a radius of about 10\,pc.
However, the clear spatial segregation between the cluster
and cloud sources suggest that the fraction of induced star formation
at early times is small. If stars formed as the HII region
expanded to its current size, we would expect dusty sources between
the cluster and cloud. There are no identifiable cores within 5\,pc
of the cluster center. However, most of the cloud lies within
the maximum radioactivity distance of 45\,pc from the cluster
and is structured into clumps with sizes and masses that match
the conditions required for enrichment in Figure~5
(Williams, Blitz, \& Stark \cite{Williams95}).

\section{The Galactic background revisited}

An essential point in the discovery of live \fe60\ in the ESS is that
its abundance is above the expected Galactic background level.
Recent $\gamma$-ray observations suggest they may not be so different,
however. The measured flux of photons at the decay energy
of an SLR can be directly converted to an abundance.
Diehl et al. (\cite{Diehl06}) show that \al26\ is produced on
a Galactic scale with an average abundance,
$[^{26}{\rm Al}]/[^{27}{\rm Al}]=8.4\times 10^{-6}$, about one sixth
the ESS value (Lee, Papanastassiou, \& Wasserburg \cite{Lee77}).
\fe60\ is less abundant and the line flux about ten times weaker
but its decay has been detected and the inferred abundance,
$[^{60}{\rm Fe}]/[^{26}{\rm Al}]=0.23$ (Harris et al. \cite{Harris05}).
This implies
$[^{60}{\rm Fe}]/[^{56}{\rm Fe}]
=[^{60}{\rm Fe}/[^{26}{\rm Al}]\times
 [^{26}{\rm Al}]/[^{27}{\rm Al}]\times
 [^{27}{\rm Al}]/[^{56}{\rm Fe}]
=0.23\times 8.4\times 10^{-6}\times 0.092=1.8\times 10^{-7}$
which is between one half to one sixth of the ESS values
(Tachibana et al. \cite{Tachibana06}; Moynier et al. \cite{Moynier05}).

Normalized by the production ratio, the observed abundances
for both \al26\ and \fe60\ are both $1.5\times 10^{-3}$.
These are, respectively, a factor of 6 and 3 higher than the
predicted ISM average (Jacobsen \cite{Jacobsen05}).
The reason for this discrepancy is not clear.
Nevertheless, it seems hard to deny the $\gamma$-ray evidence
and the implication that the observed \fe60\ background level may be
within a factor of 2 of the low end of the range of meteoritic measurements,
[\fe60]/[$^{56}$Fe]$=3\times 10^{-7}$.
Further, the Galactic star formation rate -- and therefore the
background -- may have been about two times higher at the time
the Sun was born (Rocha-Pinto et al. \cite{Rocha00}).
The match between the numbers is tantalizing and, if they hold,
the \fe60\ puzzle goes away.

Even the high end of the \fe60\ abundance measurements,
[\fe60]/[$^{56}$Fe]$=10^{-6}$, is only a factor of 3 greater than the
enhanced, early Galactic background.
Given the known inhomogeneity in the \al26\ emission
(Knodlseder et al. \cite{Knodlseder99})
might a potential explanation be that we
formed in a ``radioactive overdensity''?

The likelihood of being in an overdensity of 3 is no more than
1/3 if the new stars are uniformly distributed.
In fact, young stars congregate in spiral arms close to the supernova
that produce the background. However, the structure of the multiple
phases of the ISM is complex and not well characterized (Cox \cite{Cox05}).
The definitive answer will ultimately have to wait for higher resolution,
higher sensitivity $\gamma$-ray observations but it may also be possible
to address the issue via extragalactic observations of the ionized,
atomic, and molecular gas components
(e.g., Scoville et al. \cite{Scoville01};
Calzetti et al. \cite{Calzetti05}; Schuster et al. \cite{Schuster07}).

A crucial missing step is the journey from $\sim 10^7$\,K
hot ionized gas in supernovae remnants to $\approx 30$\,K
circumstellar disk to planetesimal. There is some evidence for
rapid cloud formation from atomic gas on Myr timescales
(Hartmann et al. \cite{Hartmann01}; Elmegreen \cite{Elmegreen00})
but there is no viable mechanism to convert the hot ionized
medium to atomic gas on comparable timescales. Note that each
delay of 1\,Myr raises the required overdensity by a factor
of 4 and 2 for \al26\ and \fe60\ respectively.

Finally, solving the \fe60\ conundrum in this way, results in the opposite
problem of enormous excesses in the abundances of the longer lived
radionucleides. This reflects the difficulty, as many studies have noted,
of matching all the abundances of all the SLR from a single source
(e.g., Harper \cite{Harper96}).

\section{Discussion and Conclusions}
I have critically evaluated several proposed mechanisms for the delivery
of \fe60\ to the ESS and found that most are highly improbable.
AGB stars have moved away from their birthsite and any cross-pollination
of a molecular core or circumstellar disk would be serendipitous.
Further, their output is too small to affect a significant fraction
of the ISM. No more than 0.1\%, and probably far less, of stars
could receive \fe60\ in this way.

Supernova may inject SLR directly into disks but the range of distances
over which disks both survive the blast and are enriched to ESS levels
is very small. Further, few disks remain by the time of the
supernova for all but the most massive, shortest-lived progenitors.
In this case, I estimate that no more than about 1\% of all protoplanetary
systems could receive \fe60\ in this way.

I was unable to assign probabilities to induced star formation
scenarios since the statistics are not well known.
The population of young stars on the boundary of HII regions appears to
be only a small fraction of that within the central cluster and the number
that form deeper in the surrounding molecular cloud.
HII regions expand rapidly and are tens of parsecs in size before
any supernova explosion. Matching the required high levels of \fe60\
places individual star forming cores well within the boundary of
mature HII regions and they would be shredded by stellar winds.
Even if a core were to collapse and form a star, the surrounding envelope
would be photoablated or swept away by the time of a supernova which
would impact, at most, a circumstellar disk. This scenario is
effectively the same as the direct disk injection hypthesis since
that used a prescription for disk evolution that included
observations of clusters with massive stars.

Larger, cluster forming clumps in the molecular cloud beyond the HII
region can capture enough supernova material and collapse fast enough
to deliver ESS levels of \fe60\ to planetary systems.
The large scale ``collect-and-collapse'' scenario for molecular
clumps neighboring a massive star forming region
may be the most promising solution to the \fe60\ puzzle.
Molecular clouds are found around many supernova remnants and,
although they may be are strongly disrupted,
star formation can occur in the surviving gas
(Huang \& Thaddeus \cite{Huang86}; Reach \& Rho \cite{Reach99};
Reynoso \& Mangum \cite{Reynoso01}).
Perhaps the Orion subgroups are an example where this happened in the
past and the Rosette cloud an example where this will happen in the future.
However, only through detailed surveys of the molecular and
protostellar content in young cloud-supernova interactions that
quantify the amount and efficiency of star formation can we assess
the likelihood that the Sun was born in such an environment.

The discrepancy between the $\gamma$-ray measurements of the 
\al26\ and \fe60\ background with model calculations needs to
be understood. Also, more sensitive cosmochemical measurements will
reduce the uncertainty on the initial abundance of \fe60\
and constrain the timing of its injection
(Bizzarro et al. \cite{Bizzarro07}, Dauphas et al. \cite{Dauphas08}).
If the high end of
of the current range is confirmed, the background can be ruled
out as the source although it may be a significant component
that reduces the amount required from a discrete source.

Finally, if there really is no common scenario that can deliver large
amounts of \fe60\
from a massive stellar core to a protoplanetary system,
we are faced with the prospect that our solar system may be
a one-in-a-hundred rarity. Astronomers have a long history of
rejecting the idea that we are special but what are the implications
in this case?

It seems reasonable to assume that however \fe60\ was delivered
to the ESS, many other SLR were incorporated in the same manner.
But if this was an unusual occurrence, then most planetary systems
have lower levels not just of \fe60\ but of other SLR,
particularly \al26. The latter is the dominant heating source
for planetesimals at early times (Hevey \& Sanders \cite{Hevey06})
and a reduced abundance would imply a different thermal history and,
potentially, result in a higher water content of terrestrial planets
(Desch \& Leshin \cite{Desch04}; Gaidos, Raymond, \& Williams \cite{Gaidos08}).

\acknowledgements
I thank Thierry Montmerle and the organizers for inviting me to
a wonderful meeting in a spectacular location. Inspiration for this
work came from stimulating conversations with Matthieu Gounelle
and Ed Young at this meeting, Jeff Hester, Steve Desch and John Bally
at a 2007 workshop, and Eric Gaidos, Sasha Krot, and Gary Huss in Hawaii.
This work is supported by the NASA Astrobiology Institute under
Cooperative Agreement No. NNA04CC08A.

\end{document}